\documentclass[aps,prb,reprint,superscriptaddress,amsmath,amssymb,floatfix]{revtex4-2}
\usepackage{graphicx}

\usepackage{hyperref}
\hypersetup{  colorlinks=true,     urlcolor=blue,     citecolor=blue,     linkcolor=blue,}
\usepackage{etoolbox}
\apptocmd{\sloppy}{\hbadness 10000\relax}{}{}

\def\l{$\lambda$-(BETS)$_2$\-GaCl$_4$}
\def\k{$\kappa$-(BETS)$_2$\-GaCl$_4$}
\def\cm{${\rm cm}^{-1}$}
\def\lb{$\lambda$-(BETS)$_2$\-GaCl$_4$}
\def\kb{$\kappa$-(BETS)$_2$\-GaCl$_4$}

\begin{document}
%\title{Charge imbalance in $\lambda$-(BETS)$_2$GaCl$_4$ and their interplay with superconductivity}
\title{Charge imbalance in \texorpdfstring{$\lambda$}{l}-(BETS)\texorpdfstring{$_2$}{2}GaCl\texorpdfstring{$_4$}{4} and their interplay with superconductivity}

\author{Olga Iakutkina}
%\author{O.~Iakutkina}
%\email{olga.Iakutkina@pi1.physik.uni-stuttgart.de}
\affiliation{1.~Physikalisches Institut, Universit{\"a}t Stuttgart, 70569 Stuttgart, Germany}

\author{Ece Uykur}
%\author{E.~Uykur}
%\email{ece.uykur@pi1.physik.uni-stuttgart.de}
\affiliation{1.~Physikalisches Institut, Universit{\"a}t Stuttgart, 70569 Stuttgart, Germany}

\author{Takuya Kobayashi}
%\author{T.~Kobayashi}
\affiliation{Graduate School of Science and Engineering, Saitama University, Saitama, 338-8570, Japan}
%\email{tkobayashi@phy.saitama-u.ac.jp}

\author{Atsushi Kawamoto}
%\author{A.~Kawamoto}
\affiliation{Department of Physics, Graduate School of Science, Hokkaido University, Sapporo 060-0810, Japan}
%\email{atkawa@phys.sci.hokudai.ac.jp}

\author{Martin Dressel}
%\author{M.~Dressel}
%\email{dressel@pi1.physik.uni-stuttgart.de}
\affiliation{1.~Physikalisches Institut, Universit{\"a}t Stuttgart, 70569 Stuttgart, Germany}

\author{Yohei Saito}
%\author{Y.~Saito}
\email{yohei.saito@pi1.physik.uni-stuttgart.de}
\affiliation{1.~Physikalisches Institut, Universit{\"a}t Stuttgart, 70569 Stuttgart, Germany}
%\affiliation{Department of Physics, Graduate School of Science, Hokkaido University, Sapporo 060-0810, Japan}

\date{\today}

\begin{abstract}
The two-dimensional organic superconductor $\lambda$-(BETS)$_2$GaCl$_4$ exhibits pronounced charge fluctuations below $T \approx 150$~K, in contrast to the sibling compound $\kappa$-(BETS)$_2$GaCl$_4$ that remains metallic down to milli-Kelvin.
Infrared spectroscopy reveals only minor splitting in the vibrational features of the latter compound,
common to other strongly dimerized $\kappa$-salts.
When the organic molecules are arranged in the $\lambda$-pattern, however, a strong vibrational $\nu_{27}(b_{1u})$ mode is present, that forms a narrow doublet.
%indicating static charge imbalance of about 2\%.
Most important, when cooling $\lambda$-(BETS)$_2$GaCl$_4$ below 150~K, two weak side modes appear
due to charge disproportionation that amounts to $2\delta=0.14e$. In analogy to the $\beta^{\prime\prime}$-type organic conductors, we propose that charge fluctuations play an important role in emerging of unconventional superconductivity in \l\ at $T_c=4.7$~K.
We discuss the possibility of a charge-density-wave that coexists with the proposed spin-density-wave state.
\end{abstract}.

\maketitle

\section{\label{sec:introduction}Introduction}
Unconventional superconductivity in strongly correlated systems such as heavy fermionic systems \cite{WHITE2015246}, transition metal oxides \cite{plakida2010high,YAMAUCHI2005874}, organic charge-transfer salts \cite{lebed2008physics,Wosnitza2019} is  a major topic in condensed matter physics.
Great interest is particularly drawn by the interplay between superconductivity and some ordered state, let it be magnetic or charge order.
Concerning the pairing mechanism in cuprates, for instance, the discussion goes about the importance of spin fluctuations, originating from the antiferromagnetic Mott insulating state, and charge fluctuations, associated with dynamical stripes \cite{doi:10.1080/000187300412248,PhysRevB.90.115137,PhysRevLett.97.097001,Torchinsky2013}.

There is a far-reaching similarity between two-dimen\-sion\-al organic conductors and cuprates \cite{McKenzie820},
including the fact that the superconducting state is found at the border of metal and Mott insulator \cite{RevModPhys.70.1039}.
The molecular systems actually constitute textbook examples of Mott insulators because the charge-transfer band is located well below the Hubbard bands split by $U \approx 0.5$-1~eV.
Commonly these salts crystallize in $D_2X$ stoichiometry, with $D$ the charge donor and $X$ the monovalent acceptor.
This implies three-quarter-filled conduction bands as long as the stacks are homogeneous; the formation of dimers, however,  results in half-filled bands.
In the first case, inter-site Coulomb repulsion $V$ governs the relevant physics, giving rise to a charge-ordered insulating ground state;
the latter case the genuine Mott system, completely characterized by the strength of $U$ with respect to the bandwidth $W$.

For half-filled Mott insulators, spin fluctuations are rather common, possibly related to superconductivity.
The $\kappa$-(BEDT-TTF)$_2$X compounds are prime examples of antiferromagnetic Mott insulators located next to superconductivity, often with some coexistence regions \cite{PhysRevB.60.3060,Kondo1998,Lefebvre00,Dressel20}.
Charge-ordered materials, on the other hand, remain non-magnetic at all temperatures; hence any effect of spin fluctuations on the superconducting ground state next to the charge-ordered phase can be rules out.
Merino and McKenzie suggested \cite{PhysRevLett.87.237002} that instead superconductivity is mediated by charge fluctuations here.
Subsequently, the importance of charge degrees of freedom for these superconducting compounds was  proven experimentally \cite{Kaiser10,*Kaiser12,PhysRevB.99.140509,*PhysRevB.99.155144}.

\begin{figure}
\centering
\includegraphics[width=1\columnwidth,clip]{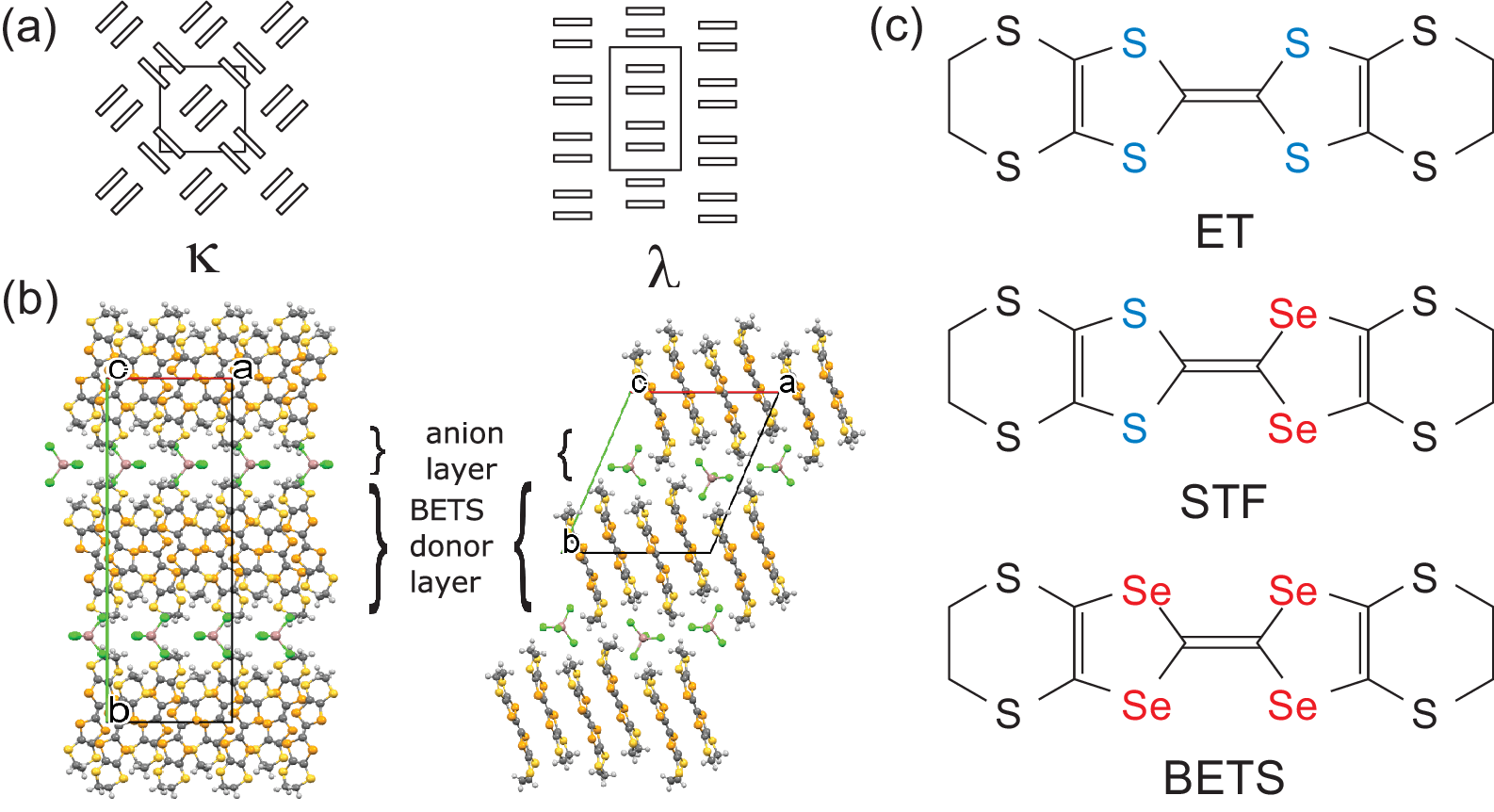}
\caption{(a)~Looking along $ac$-planes, the two-dimensional charge-transfer salts form different dimer patterns. In the $\kappa$-phase the dimers are rotated with respect to each other, while the $\lambda$-pattern is organized in stacks with two dimers, i.e.\ four molecules per unit cell.
(b)~The BETS donor layers are separated by layers of GaCl$_4^-$ anions. The organic molecules are slightly tilted in alternating directions in case of $\kappa$-BETS, while uniform in case of $\lambda$-BETS.
(c)~Most common are the donor molecules ET = BEDT-TTF, i.e. bis\-(ethylene\-di\-thio)\-tetra\-thia\-fulvalene), but sulfur can be substituted by selenium, leading to STF = BEDT-STF, and BETS = BEDT-TSF, i.e. bis\-(ethylene\-di\-thio)\-tetra\-selena\-fulvalene). The more extended selenium orbitals cause a larger bandwidth favoring metallic con\-ductivity.}
\label{fig:structure}
\end{figure}
Fig.~\ref{fig:structure}(a) illustrates that in the $\kappa$-type salts the organic molecules are arranged in distinct pairs,
while dimerization is less obvious for the $\lambda$-pattern. The interesting question now arises whether  here antiferromagnetic order is present in those compounds, how important spin fluctuations are, and whether charge fluctuations or even charge order can be observed. In this context, the series $\lambda$-$D_2$GaCl$_4$ is of particular interest, as the electronic bandwidth $W$ is varied by modifying the organic molecule $D$ =  BEDT-TTF, BEDT-STF or BETS [Fig.~\ref{fig:structure}(c)].
The strongest effect of correlations, characterized by $U/W$, occurs in $\lambda$-(BEDT-TTF)$_2$GaCl$_4$: it resembles a Mott insulator with antiferromagnetic order at $T_N=13$~K  \cite{MORI2001103,minamidate2015superconducting,Saito18}.
The intermediate compound $\lambda$-(BEDT-STF)$_2$GaCl$_4$ does not exhibit magnetic order down to lowest temperatures but rather unconventional behavior assigned to a disordered quantum spin state \cite{Saito2019disordered}.
$\lambda$-(BETS)$_2$GaCl$_4$, on the other side, remains metallic all the way down to the superconducting transition at $T_c=4.7$~K.
The picture is confirmed by pressure-dependent studies \cite{MORI2001103,minamidate2015superconducting}.

The fact that superconductivity does not occur next to the antiferromagnetically ordered phase calls for further investigations.
$^{13}$C-NMR measurements of the superconductor \l\ revealed, that the behavior of $1/T_1T$ at high temperature above 55~K is
dominated by antiferromagnetic fluctuations similar to what is observed in the $\kappa$-type salts such as $\kappa$-(BE\-DT\--TTF)$_2$\-Cu\-[N\-(CN)$_{2}$]Br.
Below 30~K, $1/T_1T$ becomes temperature independent and $\rho(T)\propto T^2$; evidencing a Fermi-liquid response.
In addition, another increase in spin-lattice relaxation rate of \l\ was observed below 10~K and interpreted as sign of Fermi surface nesting probably related to spin-density wave formation \cite{PhysRevB.96.125115}.
More light is shed on this issue by considering the series  $\lambda$-(BETS)$_2$GaBr$_x$Cl$_{4-x}$, where the Br substitution acts as negative pressure, driving the system insulating with rising $x$ \cite{tanaka1999chemical,PhysRevResearch.2.023075}.
$^{13}$C-NMR spectroscopy infers that spin degree of freedom plays an important role in the pairing mechanism of $\lambda$-type organic superconductors.

On the other hand, $^{77}$Se-NMR measurement of $\lambda$-(BETS)$_2$FeCl$_4$ and \l\ showed linewidth broadening at low temperature that is assigned to charge disproportionation \cite{Hiraki2007,hiraki2010evidence}, which can also contribute to the mechanism of superconductivity. In order to elucidate this point, comprehensive NMR investigations have been performed on \l\
probing different isotopes $^{69}$Ga and $^{71}$Ga in order to separate contributions of spin and charge dynamics \cite{PhysRevB.102.235131}.
While there is no enhancement of charge fluctuations below $T=150$~K, spin fluctuations become dominant at low temperature.
Obviously NMR spectroscopy is more sensitive to spin than to charge, calling for a detailed investigation of charge distribution in \l, for instance by probing the electronic charge directly.

For that reason we elucidate the issue of charge disproportionation by infrared spectroscopy with the focus on the range of charge sensitive molecular vibrations of \l.
We compare the findings in this paper with those on \k, which is a strongly dimerized compound that behaves metallic down to liquid-helium temperatures with no indications of strong correlations and magnetic contributions \cite{tanaka1999chemical,Pesotskii99,Pratt03}.

\section{Materials and Methods}
Single crystals of the quasi-two-dimensional organic superconductor \l\ (abbreviated as $\lambda$-BETS)
and a metal \k\ ($\kappa$-BETS hereafter)
were grown by standard electrochemical methods \cite{montgomery1994synthesis,kobayashi1993new}.
In these salts, BETS donor molecules are dimerized, forming a triangular lattice in the $\kappa$-type, and for the $\lambda$-type they are arranged on a square lattice, as illustrated in Fig.~\ref{fig:structure}(a). $\kappa$-BETS crystalizes in orthorhombic P\emph{nma} space group. Layers of donor molecules tilted in opposite directions with respect to each other, leading to a doubling of the unit cell along the $b$-axis. The unit cell contains four dimmers, and two BETS molecules in a dimer are connected by a center of symmetry [Fig.~\ref{fig:structure}(b)]. $\lambda$-BETS has lower symmetry and belong to the triclinic P$\bar1$ space group. In contrast  to $\kappa$-BETS, here the donor layers of BETS molecules are identical. As a result, it has two dimmers per unit cell as shown in Fig.~\ref{fig:structure}(b). An interesting difference between both BETS-salts is that in $\lambda$-BETS two molecules forming a dimer are not identical, but have a slightly different length of C=C double bonds.

%In these salts, BETS donor molecules are dimerized, forming a triangular lattice in the $\kappa$-type, and for the $\lambda$-type they are arranged on a square lattice, as illustrated in Fig.~\ref{fig:structure}(a). \kb\ belong to the orthorhombic P\emph{nma} space group, while \lb\ has lower symmetry, it forms the triclinic P$\bar1$ space group.

For both compounds the temperature dependence of the dc resistivity was measured along the most conducting axis using the conventional four-contacts method. The applied current was limited in the range of 100-200~$\mu$A; the resistivity data were recorded while warming up from 3 to 295~K.

Optical reflectivity measurements were performed with the light polarized perpendicular to the highly-conducting BETS planes, cf.\
Fig.~\ref{fig:structure}(b).
In this direction we can detect the molecular vibration $\nu_{27}(b_{1u})$ which is the most sensitive local probe for
the charge per molecule \cite{Dressel04,girlando2011charge,painelli1986electron}.
For this purpose, a Hyperion infrared microscope was attached to a Bruker Vertex 80v Fourier-transform infrared spectrometer.
The data were collected in the spectral range 500-8000~\cm\ between $T=295$ and 9~K. To perform the Kramers-Kronig transformation,
a constant extrapolation was used for frequencies below 500~\cm\  and the standard $\omega^{-4}$ decay was utilized as high-frequency extrapolation \cite{DresselGruner02}.

\section{Results}
\begin{figure}[b]
\centering
\includegraphics[width=1 \columnwidth,clip]{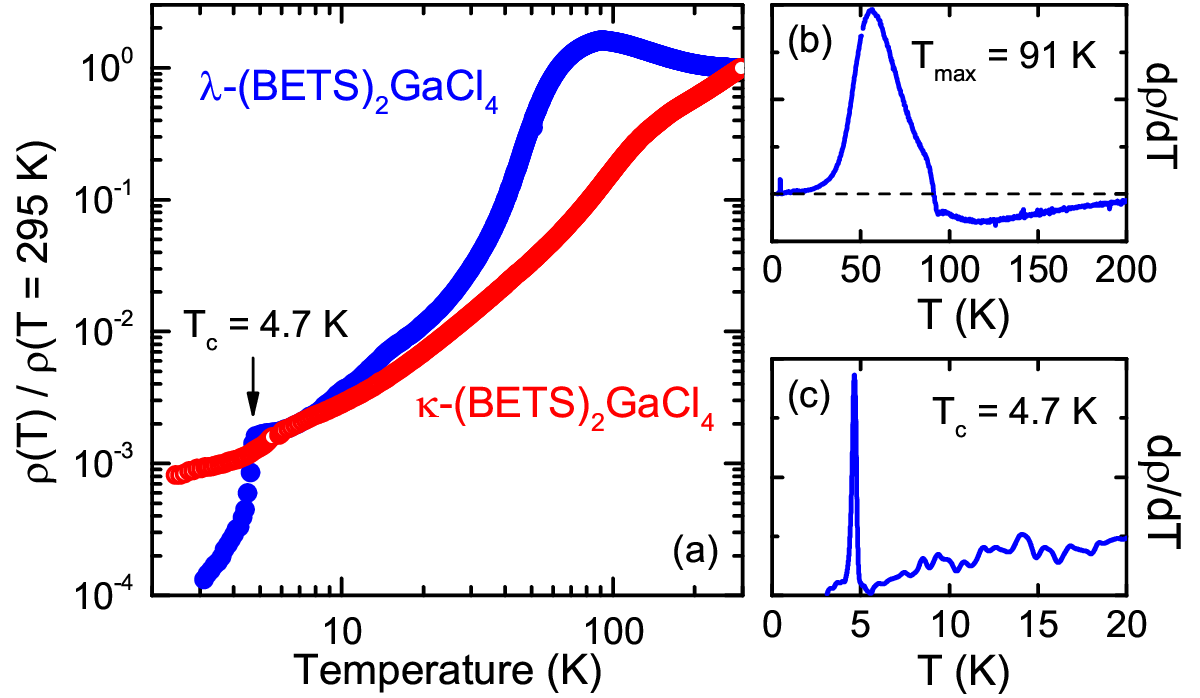}
\caption{(a) Double logarithmic plot of the relative resistivity $\rho(T)$/$\rho(295\,K)$ of \l\ (blue dots) and \k\ (red circles) measured as a function of temperature along the most conducting axes. The maximum in $\rho(T)$ for \lb\ can be best determined from the zero-crossing of the derivative d$\rho$/d$T$ at $T_{\rm max}$ plotted in panel (b) using arbitrary units. The superconducting transition occurs at $T_c = 4.7$~K as illustrated in panel (c).}
\label{fig:resistivity}
\end{figure}

In Fig.\ \ref{fig:resistivity}(a) the temperature dependence of the electrical resistivity $\rho(T)$ of \kb\ and \lb\ is plotted,
normalized to the room temperature value $\rho(T = 295 {\rm K})$.
The behavior of the two salts is rather different. \kb\ remains metallic in the entire temperature range:
$\rho(T)$  drops monotonously with minor changes in slope at elevated temperatures.
For the $\lambda$-salt a pronounced maximum is observed at $T_{\rm max} = 91$~K [Fig. \ref{fig:resistivity}(b)]; while below approximately 40~K, the $\rho(T)\propto T^2$ dependence indicates a Fermi-liquid behavior.  The compound becomes superconducting at $T_c=4.7$~K (onset at 4.8~K) as shown
in Fig.~\ref{fig:resistivity}(c). These results are consistent with previous publications \cite{kobayashi1993new}.

In order to explore possible charge distribution in organic charge-transfer salts, vibrational spectroscopy is the most sensitive and convenient method. The three modes, $\nu_{2}(a_{g})$, $\nu_{3}(a_{g})$ and $\nu_{27}(b_{1u})$, mainly involving C=C stretching,
can be utilized to determine the electronic charge per molecule.
The two totally symmetric $a_g$ modes should usually not be infrared active, but due to electron molecular-vibration (emv) coupling, they can be observed by infrared spectroscopy in the conducting plane.
As a result of the coupling to the electronic background, the position of the  $\nu_{2}(a_{g})$ and $\nu_{3}(a_{g})$ infrared feature is shifted -- compared to the Raman mode -- and depends on the  midinfrared charge-transfer band. Hence, this way a precise determination
of the charge distribution can become pretty tricky.
The antisymmetric  molecular vibration $\nu_{27}(b_{1u})$, on the other hand, is infrared active and -- in first approximation -- its frequency depends linearly on the molecular charge \cite{girlando2011charge}; this makes it perfectly suited for evaluating the charge per molecule.
Due to the stacking of the donor molecules in these salts, it can be best observed perpendicular to conducting BETS layers.
%\begin{figure}
\begin{figure}[ht]
\centering
\includegraphics[width=0.9 \columnwidth,clip]{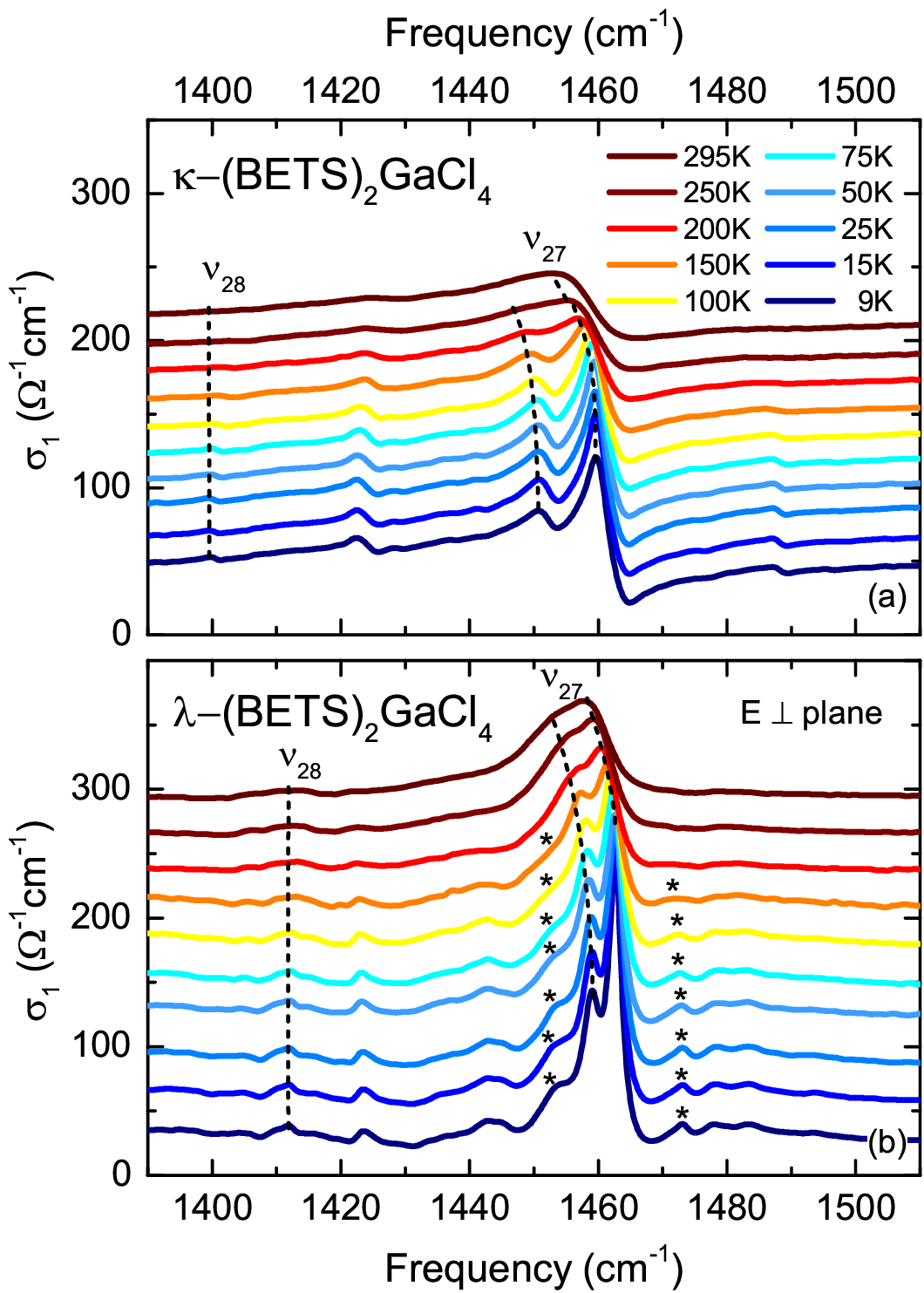}
\caption{Out-of-plane optical conductivity of (a) \kb\ and (b) \lb\ in the spectral range of the $\nu_{27}(b_{1u})$ mode
measured in the temperature range from $295$ to 9~K. For clarity reasons, the curves are shifted with respect to each other. While for \kb\ two $\nu_{27}$-related peaks are observed at low temperatures, which are gradually smeared out upon warming up;
there are two additional satellite features ($\star$) identified in the $\lambda$-salts, which appear only below approximately 150~K. }
\label{fig:v27}
\end{figure}

Fig.~\ref{fig:v27} displays the optical conductivity of $\kappa$- and \lb\ in the spectral range of the $\nu_{27}(b_{1u})$ mode, recorded at different temperatures, as indicated. Compared to the BEDT-TTF salts, this molecular vibration shows up slightly lower in frequency because four of the eight sulfur atoms are substituted by heavier selenium atoms in BETS [cf.\ Fig.~\ref{fig:structure}(c)] \cite{olejniczak:jpa-00247270}.
At room temperature, a broad peak is observed around 1453~\cm\ in \kb, which is assigned to $\nu_{27}(b_{1u})$.
When the temperature is lowered, the feature becomes narrower and below 200~K two peaks can be clearly distinguished.
The behavior found in the $\lambda$-compound is rather different: here the $\nu_{27}(b_{1u})$ mode appears as a doublet
with frequencies 1454 and 1458~\cm\ already at room temperature. Upon cooling a typical narrowing and blue shift occurs, but the separation of the two peaks does not change in the entire temperature range. We explain the presence of these two distinct peaks by the lower crystal symmetry.
Most important, however, are the two side peaks at 1451 and 1471~\cm\ that appear next to this doublet when the temperature is reduced below $T=150$~K, i.e.\ when the resistivity $\rho(T)$ starts to develop its maximum reached at $T_{\max}=91$~K (Fig. \ref{fig:resistivity}).

The $\nu_{28}(b_{1u})$ mode, which is related to ethylene-end groups, is present in both compounds. It shows up as a small peak around 1400~\cm\ in \kb, and as a doublet around 1405-1410~\cm\ in the $\lambda$-analogue due to the lower P$\bar1$ symmetry of the latter salt.
It is interesting to note that the peak seen around 1423-1425~\cm\ is very similar to one observed in the quantum spin liquid $\kappa$-(BEDT-TTF)$_2$Cu$_2$(CN)$_3$ \cite{PhysRevB.86.245103}. The tiny features near 1480~\cm\ in both compounds might correspond to $\nu_2(a_g)$. No temperature dependence was observed for these modes nor for the shoulder at 1442~\cm.

\section{Discussion}

%\subsection{Charge sensitive $\nu_{27}(b_{1u})$ mode}
\subsection{Charge sensitive \texorpdfstring{$\nu_{27}(b_{1u})$}{nu27(b1u)} mode}

\begin{figure}
\centering
\includegraphics[width=1\columnwidth, clip]{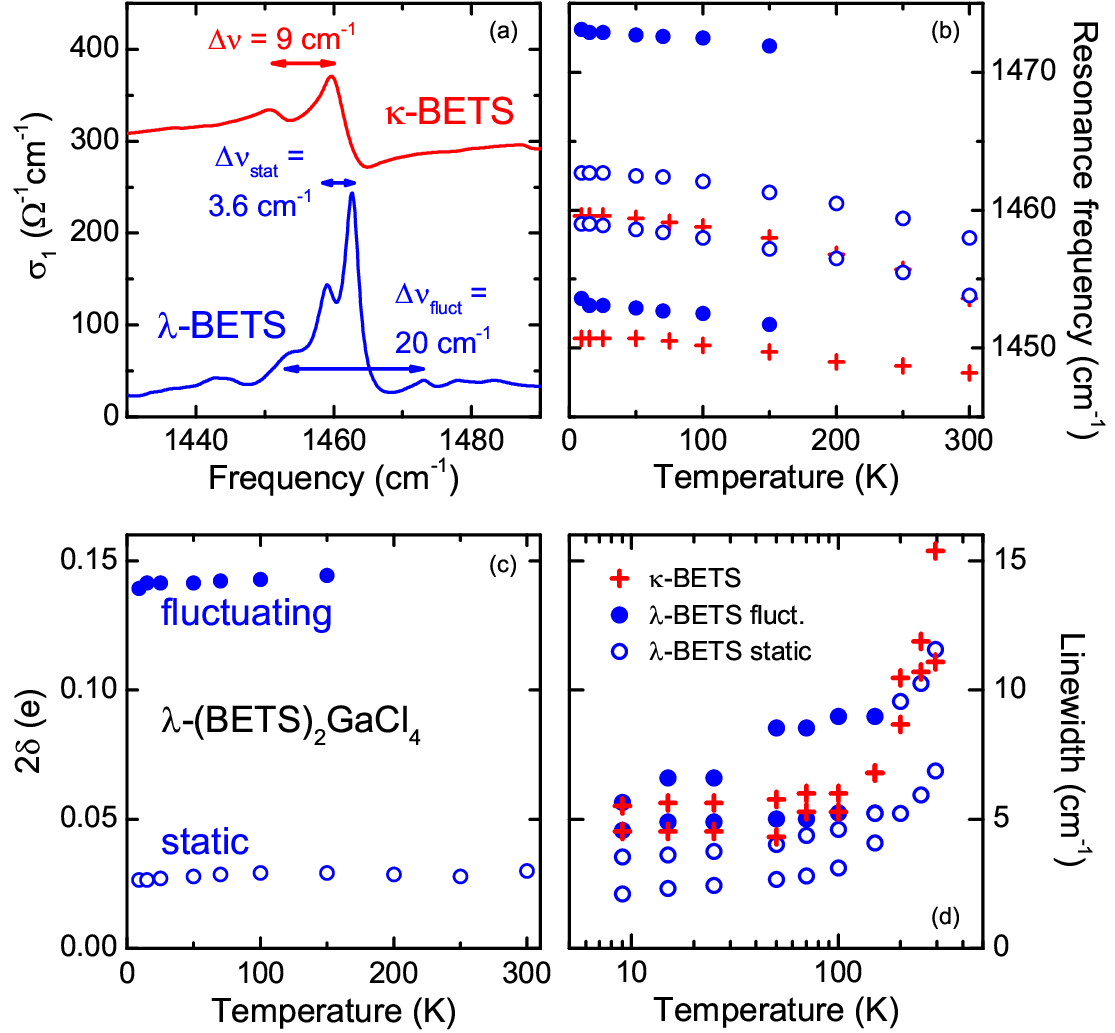}
\caption{Analysis of the vibrational feature $\nu_{27}(b_{1u})$. (a)~Comparison of the low-temperature ($T = 9$~K) optical conductivity of \kb\ and \lb\ in the range of the charge-sensitive $\nu_{27}(b_{1u})$ modes.
For \kb\ (red) a splitting of approximately 9~\cm\ occurs due to the double-layer structure.
In the case of \lb\ (blue), the pronounced doublet at 1459.1 and 1462.7~\cm\ results from crystallographically distinct molecules; the static separation is temperature independent.
In addition two broad modes appear at 1453.5 and 1473~\cm\ due to fluctuating charge.
(b)~Temperature evolution of peak positions. The red crosses correspond to the modes of \kb.
The blue solid dots represent the fluctuating modes of \lb, while the open blue circles are related to the static modes of this compound.
(c)~Temperature dependent charge imbalance $2\delta$ for \lb, calculated from the resonance frequencies of panel (b).
(d)~Linewidth of all modes plotted as a function of temperature on a logarithmic scale.}
\label{fig:fitting}
\end{figure}

Fig.\ \ref{fig:fitting}(a) illustrates the significantly different shape of the vibrational modes
in the $\kappa$- and $\lambda$-salts  by comparing the $T=9$~K spectra in the same graph.
While for \lb\ the modes can be fitted by simple Lorentz functions,
the excitation in \kb\ are surprisingly asymmetric indicating some influence of the electronic background.
Typically no sizeable conductivity occurs in the out-of-plane direction, however,
for $\kappa$-type salts the donor molecules are tilted within the plane, leading to
some appreciable coupling to the in-plane conductivity. In order to account for the interaction
of the vibrational modes with the electronic background, the Fano model
should be applied in the case of in \kb\ \cite{PhysRevB.86.245103}.

In general, a doubling of charge sensitive modes is taken as evidence for inequivalent bonds due to donor molecules containing different amount of charge $2\delta= \rho_{\rm rich}-\rho_{\rm poor}$; the observed peak separation  $\Delta \nu_{27}$  allows us to evaluate the charge imbalance $2\delta = \Delta \nu_{27} / (140~{\rm cm}^{-1}/e)$, assuming the same relation as reported for the $\nu_{27}(b_{1u})$ mode of BEDT-TTF salts \cite{girlando2011charge}. Despite the fact that the spectrum is shifted to lower frequencies by almost 10~\cm\ due to the heavier selenium atoms, our finding can be compared with the observations Sedlmeier {\it et al.} made on several $\kappa$-phase BEDT-TTF compounds \cite{PhysRevB.86.245103}.
Such minor satellite peaks occur in $\kappa$-(BE\-DT\--TTF)$_2$\-Cu\-[N\-(CN)$_{2}$]Br and $\kappa$-(BE\-DT\--TTF)$_2$\-Cu\-[N\-(CN)$_{2}$]Cl and were ascribed to Davydov-like splitting;
Maksimuk {\it et.al.}  gave a detailed description of this phenomenon including symmetry considerations and selection rules \cite{Maksimuk2001}.
The situation is similar for $\kappa$-BETS, where the unit cell is doubled
due to alternating tilting of the BETS molecules within the planes, as depinced in Fig.~\ref{fig:structure}(b).
We conclude therefore that the two maxima separated by $\Delta \nu=9$~\cm\ are due to structural
reasons. This splitting is well resolved at low temperatures but can be followed all the way up to room temperature where it is smaller by about 30\%.
We exclude the presence of static or dynamic charge disproportionation at any temperature.

The case of \lb\ is more complex because there are two rather sharp central peaks, which are distinct by only 3.6~\cm.
We assign the doublet  to the $\nu_{27}(b_{1u})$  mode of the BETS molecules bearing $+0.5~e$ charge.
Due to the low symmetry of the $\lambda$-compound, there are two inequivalent molecules in the unit cell, resulting in a doubling of this vibrational feature; the unequal bonds correspond to a static charge imbalance of only $0.02e$.
No variation with temperature is observed, except of the common thermal smearing above 100~K, demonstrated in Fig.~\ref{fig:fitting}(d).
The surprising observation is that in addition to this central doublet, two side bands appear below $T \approx 150$~K,
right when the resistivity behavior changes from metallic to semiconducting upon cooling (Fig.~\ref{fig:resistivity}).
The emergence of these features is attributed to the noticeable reorganization of charge and will be discussed in all detail
in the following.

From Figs.~\ref{fig:v27}(b) and \ref{fig:fitting}(d) we see that the width of central and side peaks in \lb\ differs significantly.
While the central modes are very narrow, with more than 5~\cm\ linewidth, the side bands are quite broad, even at low temperatures.
We suggest that the satellite  peaks are due to a dynamic charge imbalance of $2\delta_{\rm fluct}=0.14e$ emerging only for $T<150$~K. 
As a matter of fact, each satellite peak should contain a doublet, as well \footnote{There are some indications of a doublet at the high-frequency satellite, but these features are too weak to make a strong point.};
however our data do not allow such an assignment reliably due to the broad features. The splitting of the sidebands was calculated using the average frequency of the central doublet.
In addition, we should note that from $^{77}$Se-NMR measurements on \lb\
a charge distribution $2\delta = 0.15e$ was concluded considering the low-temperature Knight shift \cite{hiraki2010evidence}, rather similar to the charge disproportionation we found.
It is interesting to compare our observations with those reported for the superconductor  $\beta^{\prime\prime}$-(BEDT-TTF)$_2$\-SF$_5$\-CH$_2$CF$_2$\-SO$_3$ where at low temperature broad fluctuation modes coexist
with localized charges, seen in well separated sharp peaks \cite{PhysRevB.89.174503,*girlando2012spectroscopic}.

Finally, we want to point out that the $\nu_{27}(b_{1u})$ molecular vibrations involve the inner C=C double bonds. These should not be significantly affected by the freezing-out of the ethylene end-groups motion, that was suggested as a source of charge imbalance by Kobayashi {\it et al.} \cite{PhysRevB.102.235131}. 
Varying the cooling rate is a common approach  for investigating freezing effects. 
Yakushi {\it et al.} report that the width of the $\nu_2(a_g)$ Raman mode  is affected by the cooling rate
in $\kappa$-(BEDT-TTF)$_2$Cu$_2$(CN)$_3$ and relate that to disorder of ethylene end-groups at low temperature \cite{Foury-Leylekian18}. However, we could not identify any dependence of the charge sensitive $\nu_{27}$($b_{1u}$) infrared mode on the cooling rate \cite{Alonso2016}. 
In addition, the $\nu_{28}$ mode, which involves the terminal CH$_2$-groups, is also seen up to room temperature (Fig.~\ref{fig:v27}) with no strong variation around 100~K. 
Since in the present compound the ethylene end-groups are closely linked to the GaCl$_4^-$ anions, investigation of infrared active $\nu_3$ vibrational mode of the tetrahedral GaCl$_4^-$ anions (located in the far-infrared range) is highly appreciable to shed light on the influence of ethylene motion on physical properties of \lb.
More specific studies in this regard are in progress.

As far as the mode strength is concerned, the intensity of vibrational modes
generally increases with electronic charge, corresponding to a shift to lower frequencies,
as shown by quantum mechanical calculations \cite{girlando2011charge,*girlando2012spectroscopic}.
Indeed, the peak observed in \lb\ at $1453$~\cm\ has a much higher intensity compared to one at $1573$~\cm.

We have seen in Fig.~\ref{fig:resistivity}
that the $\kappa$-phase salt is a metal at all temperatures; therefore it is not surprising to find only
broad vibrational modes even in the perpendicular direction.
The non-monotonous resistivity $\rho(T)$ for \lb\ that goes through a maximum around $T_{\rm max} = 91$~K upon cooling, indicates that electronic charges are prone to localization, because electronic correlation are more important \cite{Pesotskii99}.
The presence of superconductivity right at the metal-insulator transition in the phase diagram raises the question about possible mechanisms.

Now we want to draw the attention to the series of $\beta^{\prime\prime}$-(BEDT-TTF)$_2$$X$ salts, that was subject to extended investigations for many years \cite{Ward00,*Schlueter01,PhysRevB.89.174503,Dressel20}, including comprehensive optical studies that focussed on the interplay of charge order and superconductivity. By variation of the all-organic anions $X$, completely insulating, charge-ordered, superconducting and metallic systems can be achieved.
In the latter compound, $\beta^{\prime\prime}$-(BEDT-TTF)$_2$\-SF$_5$\-CHFCF$_2$\-SO$_3$, only a very low degree of charge fluctuations is
seen, resulting a very broad feature in the optical conductivity. The superconductor $\beta^{\prime\prime}$-(BEDT-TTF)$_2$\-SF$_5$\-CH$_2$CF$_2$\-SO$_3$,
on the other hand, exhibits well distinct peaks corresponding to fluctuating charges \cite{Kaiser10,PhysRevB.99.140509,*PhysRevB.99.155144}; very much fostering the idea of charge fluctuations as glue for superconductivity.

For a more quantitative analysis, the vibrational features were fitted by one Fano function each in case of $\kappa$-type;
for the $\lambda$-type salt one Lorentzian each suffices . The temperature evolution of the peak positions and the calculated charge imbalance $2\delta$ is plotted in Fig.\ref{fig:fitting}(b) and (c), respectively. It is surprising that $2\delta$ does not change significantly upon cooling for any of the features. This behavior is completely different to quasi-one-dimensional (TMTTF)$_2$$X$ compounds where the charge imbalance increases below $T_{\rm CO}$ in a mean-field manner \cite{dressel2012comprehensive,knoblauch2012charge,rosslhuber2018structural,PhysRevB.94.195125}.
The present behavior instead resembles the finding in the superconducting $\beta^{\prime\prime}$-(BEDT-TTF)$_2$SF$_5$CH$_2$CF$_2$SO$_3$ and metallic $\beta^{\prime\prime}$-(BEDT-TTF)$_2$SF$_5$CHFSO$_3$ \cite{PhysRevB.99.155144}.

The temperature dependence of the linewidth is displayed in Fig.\ref{fig:fitting}(d) for all modes considered. The peaks of the $\kappa$-salt are mainly affected by the electronic background and thermal broadening. A  similar behavior is observed for the narrow static modes in \lb. The temperature evolution of the sidebands is much smaller, they remain rather broad even at low temperatures. In the following we elaborate on the possible origin of charge fluctuations below $T=150$~K.

\subsection{Interplay of broken symmetry ground states in spin and charge sectors}

In $\lambda$-type salts the Fermi surface comprises cylindrical portions with one-dimensional flat sections \cite{kobayashi1996new,mielke2001superconducting}, thus nesting effects and possible instabilities are expected. Indeed, the spin-density-wave (SDW) state accompanied by the metal-insulator transition was recently observed in $\lambda$-(BETS)$_2$GaBr$_x$Cl$_{4-x}$ series with $x=0.75$ \cite{PhysRevResearch.2.023075}. No doubt that this state of itinerant antiferromagnetism certainly affects electronic properties of adjacent \lb, where the increase of $1/T_1T$ just above the superconducting transition was ascribed to the SDW fluctuations \cite{PhysRevB.96.125115}, and as a result, magnetic fluctuations were considered as a pairing mechanism of superconductivity.

Our current infrared  measurements reveal that in \lb\ not only magnetic fluctuations but also charge fluctuations are present at low temperature.
In this case, the simple Hubbard model is insufficient to describe the underlying physics, instead both on-site Coulomb repulsion $U$ and inter-site interaction $V$ has to be takes into account.
The properties of a square lattice at half-filling by using the extended Hubbard model,
infer coexistence of spin and charge density waves (SDW and CDW) near the superconducting state when $V=U/4$ \cite{PhysRevB.70.094523}. Along this lines, we may now explain how superconductivity arises when both charge- and spin-fluctuations are present.

For the one-dimensional Bechgaard salt (TMTSF)$_2$\-PF$_6$, for instance,
several groups have established the low-temperature SDW phase next superconductivity \cite{Vuletic02,PhysRevLett.110.167001,Gerasimenko14,Narayanan14}. X-ray scattering experiments
revealed features associated with a CDW state even for temperatures below $T_{\rm SDW}$ \cite{pouget1996structural,pouget1997x};
thus Clay, Mazumdar and collaborators concluded that charge and spin density waves coexist \cite{PhysRevLett.82.1522,Clay19}.
Since for (TMTSF)$_2$PF$_6$ and $\lambda$-(BETS)$_2$GaBr$_{0.75}$Cl$_{3.25}$
a qualitatively similar situation is observed, we expect that in the latter compound a CDW state can coexist with a SDW ground state.
This implies that charge fluctuations and spin fluctuations may occur next to and even cause the superconducting state. Interestingly, several of the strongly dimerized $\kappa$-type salts
with dominant magnetic fluctuations exhibit superconducting transition temperatures of 10 K and more,
while those systems with no or weak dimerization $T_c$ barely  exceeds 5~K. We might speculate that in the present case of $\lambda$-BETS the charge imbalance prevents a larger $T_c$.
Further investigations of the neighboring insulating state are required to eventually prove this suggestion. Pressure-dependent studies could be an option to tune the systems continuously albeit rather challenging. Beside a few examples in the field of quantum spin liquids \cite{Li19,Pustogow21,*Rosslhuber21}, here
we want to mention the dimer Mott insulator $\beta^{\prime}$-(BEDT-TTF)$_2$ICl$_2$ where superconductivity under pressure is discussed near spin- and charge-density-wave phases \cite{Hashimoto15}.
%$\beta^{\prime}$-EtMe$_3$Sb[Pd(dmit)$_2$]$_2$ \cite{Li19}, $\kappa$-(BEDT-TTF)$_{2}$Cu$_{2}$(CN)$_{3}$ \cite{Rosslhuber21}.

Finally, let us recall the series $\alpha$-(BEDT-TTF)$_2$\-$M$Hg(SCN)$_4$, with $M$ = K, Tl, Rb, NH$_4$,
which is a typical quarter-filled system with clear indications of charge order.
Interestingly, $M$ = NH$_4$ is superconducting while for the other compounds various density-wave scenarios have been suggested because the electronic structure contains quasi-one-dimensional bands at $E_F$ \cite{Mori1990,Dressel04,Dressel03,*Dressel05,*Drichko06,*Merino06,Kartsovnik93,*Christ00,*Andres05,Kawai08}.
As the system is at the verge of a charge-ordered phase, the effect of charge fluctuations on superconductivity is evident and should become subject of further investigations, including the recently synthesized modified compounds which also show density wave states \cite{Ohnuma18,*Ohnuma19}.

At this point, we see that both charge and spin fluctuations may be present at low temperature in \lb,
but their influence on superconductivity is not completely clear.
In the case of the half-filled $\kappa$-(BEDT-TTF)$_2$X salts, where $V=0$ and only $U$ is present,
charge fluctuations were found to be insignificant.
For the quarter-filled $\beta^{\prime\prime}$-type systems, where $V$ is strong,
a link between charge fluctuations and superconductivity were suggested theoretically and proven experimentally \cite{PhysRevLett.87.237002,Kaiser10,PhysRevB.99.140509}.
This is consistent with the phase diagram proposed by Onari {\it et al.} \cite{PhysRevB.70.094523}. When both $U$ and $V$ are present,
it is difficult to say whether spin- and charge- fluctuations compete, whether they enhance each other,
and in which way their interplay leads to superconductivity.  Further investigations are needed to draw firm conclusions.
Nevertheless, the present study clearly shows that when going from the bare metal \kb\ to the superconductor \lb, the influence of the charge degrees of freedom increases.
The $\lambda$-type salts are another example for the idea that charge fluctuations
affect superconductivity.

\section{Conclusions}

In summary, infrared optical spectroscopy was carried out in the range of the charge-sensitive $\nu_{27}(b_{1u})$ mode in order to learn about the interplay of charge order and superconductivity in the two-dimensional organic conductor \lb, using \kb\ as a reference.
In the latter compound, alternating layers and metallic conductivity results in Fano-shaped slightly distinct modes. When we go to the superconducting \lb,
we observe slight splitting of the $\nu_{27}$ mode due to the lower symmetry structure.
Most important, two side modes appear upon cooling which are assigned to a dynamical charge imbalance of $2\delta \approx 0.14e$. We suggest a situation similar to $\alpha$- and $\beta''$-type salts, where charge fluctuations are closely linked to superconductivity and contribute to their pairing mechanism.
In accord with theoretical proposals, we suggest that these charge fluctuations originate in a dynamic charge-density-wave state,
which coexists with the dynamic spin-density-wave state evidenced by NMR measurements.
A complete picture should eventually explain the properties of charge-ordered $\alpha$- $\beta^{\prime}$- and $\beta^{\prime\prime}$-compounds together with the observations of \lb.

\section{Acknowledgements}
%We acknowledge valuable comments by Dr.\ Yohei Saito who initiated this research project in Stuttgart.
We would like to thank K. Tsuji and also G.  Untereiner  for  her crucial  technical  support.
The work was supported by the Deutsche Forschungsgemeinschaft via DR228/\mbox{39-3}.

\end{document}